\documentclass[aip,jcp,10pt,twocolumn,superscriptaddress,a4paper,nofootinbib]{revtex4-1}
\usepackage{amsmath}
\usepackage{amsfonts}
\usepackage{amssymb}
\usepackage{graphicx}

\usepackage{color}
\newcommand{\bra}[1]{\langle #1|}
\newcommand{\ket}[1]{|#1\rangle}
\newcommand{\braket}[2]{\langle #1|#2\rangle}

\begin{document}
\title{Embedded-Cluster Calculations in a Numeric Atomic Orbital Density-Functional Theory Framework}

\author{Daniel Berger}
\email{daniel.berger@ch.tum.de}
\affiliation{Chair for Theoretical Chemistry and Catalysis Research Center, Technische Universit{\"a}t M{\"u}nchen,\\ Lichtenbergstr. 4, D-85747 Garching, Germany}

\author{Andrew J. Logsdail}
\email{a.logsdail@ucl.ac.uk}
\affiliation{University College London, Kathleen Lonsdale Materials Chemistry, Department of Chemistry, 20 Gordon St., London, WC1H 0AJ, UK}

\author{Harald Oberhofer}
\affiliation{Chair for Theoretical Chemistry and Catalysis Research Center, Technische Universit{\"a}t M{\"u}nchen,\\ Lichtenbergstr. 4, D-85747 Garching, Germany}

\author{Matthew R. Farrow}
\affiliation{University College London, Kathleen Lonsdale Materials Chemistry, Department of Chemistry, 20 Gordon St., London, WC1H 0AJ, UK}

\author{C. Richard A. Catlow}
\affiliation{University College London, Kathleen Lonsdale Materials Chemistry, Department of Chemistry, 20 Gordon St., London, WC1H 0AJ, UK}

\author{Paul Sherwood}
\affiliation{Scientific Computing Department, STFC Daresbury Laboratory, Daresbury, Warrington, UK}

\author{Alexey A. Sokol}
\affiliation{University College London, Kathleen Lonsdale Materials Chemistry, Department of Chemistry, 20 Gordon St., London, WC1H 0AJ, UK}

\author{Volker Blum}
\affiliation{Department of Mechanical Engineering and Materials Science, Duke University, Durham, NC 27708, USA}

\author{Karsten Reuter}
\affiliation{Chair for Theoretical Chemistry and Catalysis Research Center, Technische Universit{\"a}t M{\"u}nchen,\\ Lichtenbergstr. 4, D-85747 Garching, Germany}

\begin{abstract}
We integrate the all-electron electronic structure code {\tt FHI-aims} into the general {\tt ChemShell} package for solid-state embedding (QM/MM) calculations. A major undertaking in this integration is the implementation of pseudopotential functionality into {\tt FHI-aims} to describe cations at the QM/MM boundary through effective core potentials and therewith prevent spurious overpolarization of the electronic density. Based on numeric atomic orbital basis sets, {\tt FHI-aims} offers particularly efficient access to exact exchange and second order perturbation theory, rendering the established QM/MM setup an ideal tool for hybrid and double-hybrid level DFT calculations of solid systems. We illustrate this capability by calculating the reduction potential of Fe in the Fe-substituted ZSM-5 zeolitic framework and the reaction energy profile for (photo-)catalytic water oxidation at TiO$_2$(110).
\end{abstract}


\maketitle

\section{Introduction}

Periodic boundary conditions (PBC) and their use in supercell geometries constitute a most efficient approach to bulk and surface electronic structure calculations of solids. While focusing the computational effort on the finite supercell, this approach elegantly captures e.g. long-range electrostatic effects and in particular the delocalized electronic bonding in metallic systems. Much of this elegance and efficiency is lost though in the application to localized perturbations of the lattice periodicity, for instance in form of defects or adsorbates. Large supercells are then needed to suppress spurious interactions with periodic images, and for explicitly charged systems intricate correction schemes are required to reach convergence with respect to supercell size at all \cite{MakovPayne1995,LanyZunger2009,Freysoldt2009}. For non-metallic systems this has long motivated an alternative solid-state embedding approach in form of hybrid quantum and molecular mechanical (QM/MM) calculations \cite{Bernstein2009}. 
Here, a finite ``quantum region'' is embedded in a surrounding environment, generally modelled using classical molecular mechanics (``MM region''). Such aperiodic embedded-cluster models do not suffer from spurious interactions between defects, adsorbates and charges. Simultaneously, the extended MM environment ensures correct long-range electrostatics and elasticity, and mitigates quantum confinement effects in comparison to approaches based on bare or hydrogen-saturated clusters. 

The exploration of advanced density-functional theory (DFT) exchange-correlation (xc) functionals for solid-state applications adds another motivation for the QM/MM ansatz. With computationally particularly efficient implementations presently achieved for localized basis sets and finite systems, solid-state embedding promises unprecedented access to extended systems. With this objective, we here present a QM/MM implementation integrating the {\tt FHI-aims} program package \cite{Blum2008} into the general {\tt ChemShell} framework \cite{Sherwood2003,Sokol2004}. Exploiting tailored numeric atomic orbital (NAO) basis sets\cite{Blum2008,IgorZhang2013}, the resolution-of-identity technique \cite{Ren2012} and massively parallel algorithms \cite{Havu2009,Auckenthaler2011,ELPA_article} {\tt FHI-aims} offers a wealth of tractable functionality beyond semi-local DFT. This includes in particular exact exchange and second order perturbation theory (PT2)\cite{Ren2013JMS}, rendering {\tt FHI-aims} an ideal platform to 
test the 
performance of hybrid and double hybrid xc functionals. On the other side, {\tt ChemShell} is a powerful computational chemistry environment, which takes over the communication and data handling in hybrid QM/MM calculations. For solid state embedding in particular, Chemshell allows the use of polarizable core-shell model potentials to fit the electrostatic potential to that of a periodic calculation \cite{Sherwood2003,Sokol2004}.

In order to prevent spurious charge leakage out of the QM and into the MM region, solid-state embedding often requires the introduction of an intermediate shell at the QM boundary. In this shell all cations are replaced by effective core potentials, which prevent the spurious overpolarization of the electron density through adjacent bare positive MM charges, i.e. a spurious charge leakage into the MM region. A major part of the {\tt FHI-aims}--{\tt ChemShell} coupling therefore concerns the implementation of efficient pseudopotential (PP) functionality into the all-electron full-potential code {\tt FHI-aims}. Tailored to the NAO basis sets of {\tt FHI-aims} we specifically achieve this through norm-conserving PPs of Kleinman-Bylander type \cite{KleinmanBylander1982}.

The performance and validity of the resulting QM/MM implementation is demonstrated using Fe-substituted ZSM-5 zeolites and the TiO$_2$(110) surface as examples
for semi-covalent and ionic systems, respectively. For the former, the purely electrostatic embedding is successfully benchmarked against previous QM/MM calculations using the {\tt Gamess-UK} package for the QM calculations \cite{UCL.23}. For the latter, we exploit the {\tt FHI-aims} capability of calculating finite and periodic systems within the same numerical framework and systematically compare the solid-state embedding against PBC slab calculations. Both for reduction potentials of Fe in Fe-substituted ZSM-5 and adsorption energies of reaction intermediates in the (photo-)catalytic water oxidation at TiO$_2$(110) the obtained results underscore the importance of capturing the long-range electrostatics of the extended systems and the need to scrutinize more readily available semi-local DFT energetics.

\section{Theory} 

\subsection{Solid-state embedding with {\tt ChemShell}}

\begin{figure}
\includegraphics[width=8.5 cm]{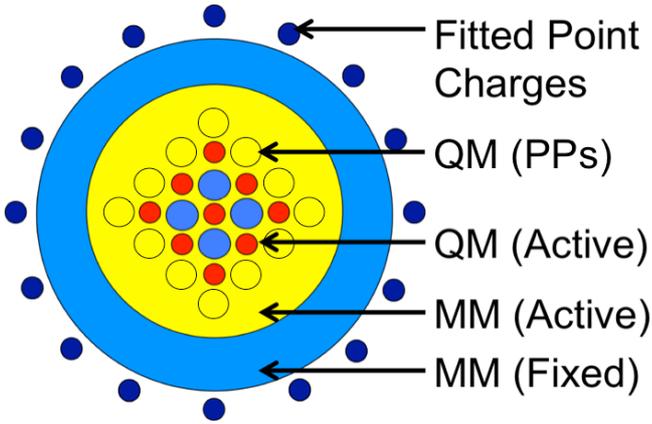}
\caption{Schematic representation of the embedded-cluster models used in the QM/MM calculations: The central QM atoms, represented by individual red and blue circles, is surrounded by regions of active (yellow) and frozen (light blue) MM atoms. If necessary, such as for ionic systems, an additional embedding potential (yellow circles) is included around the QM region in the form of atom-centred pseudopotenials (PPs). The outer shell of point charges (dark blue) is fitted to ensure reproduction of the correct long-range electrostatic potential within the active region.}
\label{fig1}
\end{figure}
 
The embedded-cluster models used in the QM/MM {\tt ChemShell} calculations generally consist of multiple concentric regions, each of which is described using more approximate methods as one moves further from the center of the cluster, as illustrated in Figure \ref{fig1}. The central region of the embedded cluster is described quantum mechanically, with the applied level of theory being DFT, Hartree-Fock (HF) or higher post-HF as offered by the employed QM calculator. Atoms in the outer MM regions are treated at the level of interatomic potentials and are represented within the QM calculations as external point charges. The MM region is typically divided into an inner ``active'' part, where atoms are allowed to relax their positions, and an outer part where atoms are constrained to their lattice positions. 
Shell model potentials can be employed in the MM region for an accurate treatment of polarizable materials.
A final exterior shell of point charges is added to the embedded cluster, with the point charges fitted to reproduce the full electrostatic embedding potential of an infinite bulk reference calculation within the regions of physical and/or chemical interest (the ``active'' region) \cite{Sokol2004}.

This zero-reference potential for a given extended "perfect" system can be e.g. 
the bulk of a three-dimensional periodic solid, i.e. a crystal, or one of its surfaces or interfaces with another material or solvent. In this way a common absolute potential reference is provided for all embedded-cluster calculations. The latter is of special importance when modelling atomic and electronic processes involving a change in the charge (oxidation) state or comparing systems in different charge states. For example, calculation of defect levels in semiconductors and insulators that represent energies of defect ionisation with electrons or holes becoming delocalised over conduction states depend crucially on this property. Another example is given by complementary oxidation-reduction cycles in acid-base and redox chemistry in solutions that involve ions or electrons to be removed or added to the system from remote locations. In all of these cases, a straightforward summation of the electrostatic contributions to the 
potential over a neutral extended system is typically performed using Ewald-like techniques \cite{Sokol2004}. The values of this potential on a grid of points spanning the active region serve then as reference for the embedded cluster model, i.e. the values of the point charges around the embedded cluster are fitted to reproduce optimally 
the potential of the extended reference calculation.

The energy of the QM region, $E^{\rm QM}$, thus combines the actual QM Hamiltonian ($\hat{H}^{\rm QM}$) with the surrounding environment:
\begin{equation}
E^{\mathrm{QM}} = \bra{\Psi} \hat{H}^{\mathrm{QM}} + \hat{V}^{\mathrm{MM}} \ket{\Psi} + V_{\mathrm{N}}^{\mathrm{MM}} \quad ,
\label{eq1}
\end{equation}
where $\hat{V}^{\mathrm{MM}}$ is the external embedding potential acting on $\ket{\Psi}$ from the surrounding point charges, and $V_{\mathrm{N}}^{\mathrm{MM}}$ is the Coloumbic interaction between the QM nuclei and the MM charges. For covalently bound, or semi-covalent/semi-ionic systems such as siliceous frameworks, saturation of dangling bonds on the exterior of the QM region with ``link''-atoms is a commonly used methodology \cite{UCL.166, UCL.167}. Such a setup is sufficient for strongly covalent materials with directional bonds, for example, of a $\sigma$ character. However, in the case of heteropolar semiconductors and ionic insulators, direct linkage of the field of MM point charges with the QM region as in eq. (\ref{eq1}) results in a spurious overpolarization of the wavefunction, aka a charge leakage towards the attractive cations of the immediately surrounding MM region. In this instance, the introduction of a boundary region is necessary between QM and MM atoms. In the present work this 
is 
realized by replacing the cationic point charges in an intermediate shell surrounding the QM region with norm-conserving PPs of Kleinman-Bylander type mimicking the corresponding ions. The energy of the QM region then reads:
\begin{equation}
E^{\mathrm{QM}} = \bra{\Psi} \hat{H}^{\mathrm{QM}} + \hat{V}^{\mathrm{PP}} + \hat{V}^{\mathrm{MM}} \ket{\Psi} + V_{\mathrm{N}}^{\mathrm{MM}} \quad ,
\label{eq2}
\end{equation}
where $\hat{V}^{\mathrm{PP}}$ is the potential acting on $\ket{\Psi}$ in form of the PPs.

The MM energy term ($E^{\mathrm{MM}}$), discussed in detail previously \cite{Sokol2004}, depends on the problem being addressed, with the {\tt ChemShell} package accommodating various forms: the General Utility Lattice Program ({\tt GULP}) offers comprehensive MM functionality, especially in the implementations for strongly polarisable materials such as ionic oxides \cite{GULP}; an alternative such as {\tt DL\_POLY} provides excellent parallelisation of larger molecular dynamics problems, and contains forcefield implementations more suited for semi-covalent systems \cite{UCL.170}. The total energy ($E^{\mathrm{tot}}$) of the embedded-cluster system is finally given by simple additive definition:
\begin{equation}
E^{\mathrm{tot}} = E^{\mathrm{QM}} + E^{\mathrm{MM}} \quad .
\end{equation}

{\tt ChemShell} manages the properties of an embedded cluster during calculations, outsourcing the calculation of energy and force terms through the varied QM and MM interfaces. The program couples a number of quantum chemistry and classical forcefield software packages: In a given geometry the QM driver computes the energy as given in eqs. (\ref{eq1}) or (\ref{eq2}), and the forces acting on all particles (atoms, PPs and point charges). The forcefield software then evaluates further forces on point charges, and geometry optimisation is performed using the {\tt DL-FIND} routine \cite{UCL.103}. Iteratively, this leads to self-consistent embedding and polarization. In this work we use the {\tt FHI-aims} package for the QM calculations and modified it to be run as a library package rather than runtime executable. For the MM calculations {\tt GULP} and {\tt DL\_POLY} are used for the ionic and semi-covalently bound showcase systems, respectively.

\subsection{{\tt FHI-aims} as QM calculator}

{\tt FHI-aims} is a full-potential all-electron electronic structure theory package providing both DFT and "beyond-DFT" functionality \cite{Blum2008,Ren2012,Ren2013JMS}. Notably, this comprises efficient treatments beyond standard semi-local DFT, such as bare or screened single-determinant exchange; quantum-chemical perturbation theory for the Coulomb interaction, e.g., second-order M{\o}ller-Plesset (MP2) perturbation theory, bare (MP2) or screened (GW) self-energies for single-electron excitation energies; or the random-phase approximation (RPA) in the adiabatic connection fluctuation dissipation theorem. It is based oni hierarchical sets of all-electron atom-centered NAO basis functions of the form
\begin{equation}
\braket{{\bf r}}{\phi_\alpha} = \phi_\alpha({\bf r}) =  \frac{u_l(r)}{r} Y_{lm}(\Omega) \quad ,
\end{equation}
where $Y_{lm}(\Omega)$ are the spherical harmonics, and the radial functions $u_l(r)$ are numerically tabulated and therefore fully flexible. Each basis function is strictly localized inside a given radius, which enables a highly efficient computation of both finite and (within PBCs) infinite systems within the same numerical framework. The regular basis set levels (called tiers, i.e. tier1, tier2, $\ldots$), as well as the recent valence-correlation consistent NAO-VCC-nZ basis sets for light elements \cite{IgorZhang2013} are constructed to enable systematic accuracy improvements from fast qualitative to meV per atom. 
Radial functions are integrated on a dense one-dimensional grid with logarithmic spacing. Global integration is efficiently achieved on a sparser three-dimensional concatenation of atom-centered grids \cite{Becke1988,Delley1990} (see ref. \onlinecite{Havu2009} for details of the present implementation). Converting between logarthimic and global integration grid is done by using cubic splines. 
When working with embedding potentials (particularly, bare monopoles) inside the range of other integrands (e.g., the Kohn-Sham Hamiltonian matrix elements), these singularities must be integrated 
accurately, e.g., by placing an additional atom-centered grid on the site of the embedding potential.

With analytic forces provided, the major effort in the interfacing of {\tt FHI-aims} with the {\tt ChemShell} environment is the implementation of PP-functionality to describe cations at the boundary between the QM and MM region. For numerical efficiency reasons described below, we opt for normconserving PPs of Kleinman-Bylander type \cite{KleinmanBylander1982}. In general, PPs are constructed to replace the atomic all-electron potential such that core states are eliminated and the then missing orthogonalization constraint of valence and core wavefunctions is achieved through appropriate scattering properties of the PP \cite{Austin1962,Fuchs1999}. In a real-space representation convenient to use with localized basis sets, this energy-dependent scattering is expressed by a dependence on the angular momentum,
\begin{equation}
\hat{V}^{\rm PP} = \sum_{lm} \ket{Y_{lm}} V_{l}(r) \bra{Y_{lm}} \quad .
\end{equation}
For every angular momentum $l$ up to a maximum $l_{\rm max}$, usually 2 or 3, there is thus a different spherically-symmetric PP channel $V_l(r)$ with corresponding eigenfunctions
\begin{equation}
\braket{{\bf r}}{\psi_{lm}} = \psi_{lm}({\bf r}) = \frac{v_l(r)}{r} Y_{lm}(\Omega) \quad .
\label{equation:psi}
\end{equation}
The radial functions $v_l(r)$ as well as the potentials $V_{l}(r)$ are commonly tabulated on a logarithmic grid. Our implementation is specifically tailored to read the tabulated format of the *.cpi PP files provided with the {\tt FHI98PP} package \cite{Fuchs1999, FHI98PP}. In principle, any kind of PP can be used though, as long as it is made available in the *.cpi format of {\tt FHI98PP}. An almost complete database of PPs is e.g. available on the {\tt abinit} webpage \cite{abinit}.

Mimicking an ion with a net charge, all PP channels $V_l(r)$ must embody the same Coulomb behaviour and are thus independent of the angular momentum in the far field, i.e. all $V_l(r)$ are the same outside the core radius $r_{\rm core}$ of the PP by construction. One channel can thus be chosen to embody all ionic long-range behavior. This local potential $V^{\rm loc}(r)$ then acts on all electrons independent of their angular momentum, while the remaining channels $\delta V_l = V_l(r) - V^{\rm loc}(r)$ are now short-ranged with $\delta V_l(r) = 0$ for $r > r_{\rm core}$. Mathematically, the choice of the local component is hereby arbitrary. In practice a proper choice is crucial for the performance and transferability of the PP \cite{Gonze1991}. For instance, the routine {\tt pswatch} in the {\tt FHI98PP} package provides full analysis capabilities according to rigorous criteria by Gonze \textit{et al.} \cite{Gonze1991}.

Numerically, it is most convenient to further transform the PP into the fully separable form of Kleinman-Bylander, where the short-ranged part $\delta V^{\rm KB}$ is a fully nonlocal operator in ${\bf r}$-space \cite{KleinmanBylander1982}
\begin{eqnarray}
\bra{{\bf r}} \hat{V}^{\rm PP} \ket{{\bf r'}} &=& \bra{{\bf r}} \hat{V}^{\rm loc} \ket{{\bf r'}} + \bra{{\bf r}} \hat{V}^{\rm KB} \ket{{\bf r'}} \\ \nonumber
=\; V^{\rm loc}(r) \delta({{\bf r} - {\bf r'}}) &+& \sum_{l=0}^{l_{\rm max}} \sum_{m=-l}^{l} \braket{{\bf r}}{\chi_{lm}} E_l^{\rm KB} \braket{\chi_{lm}}{{\bf r'}} \quad .
\end{eqnarray}
Here, the projector functions $\ket{\chi_{lm}}$ are defined as 
\begin{equation}
\ket{\chi_{lm}} = \frac{1}{\sqrt{\bra{\psi_{lm}}(\delta V_l)^2 \ket{\psi_{lm}}}}\delta V_l \ket{\psi_{lm}}
\end{equation}
with energies 
\begin{equation}
E_l^{\rm KB} = \frac{\bra{\psi_{lm}}(\delta V_l)^2 \ket{\psi_{lm}}}{\bra{\psi_{lm}}\delta V_l\ket{\psi_{lm}}} \quad .
\end{equation}

The local potential part is straightforward to implement. It is mapped from the (finite) logarithmic grid around the PP center onto the global integration grid with the help of cubic splines \cite{Delley1990,Blum2008}, and its long-range part beyond the limits of the atom-centered grid is extrapolated by its Coulombic behavior (formal charge over distance). In the evaluation of the Hamiltonian matrix, $V^{\rm loc}$ is then evaluated exactly like all other local potentials. The general matrix expression of the nonlocal potential involves the evaluation of a sum over the projections of every basis function $\ket{\phi_\alpha}$ on to every Kleinman-Bylander projector function $\ket{\chi_{lm}}$
\begin{equation}
\label{formular:nonlocal}
\bra{\phi_{\alpha}}\hat{V}^{\rm KB}\ket{\phi_{\beta}} = \sum_{l=0}^{l_{\rm max}} \sum_{m=-l}^{l} \braket{\phi_{\alpha}}{\chi_{lm}} E_l^{\rm KB} \braket{\chi_{lm}}{\phi_{\beta}} \quad .
\end{equation}
For a given geometry, this matrix is computed once and is added to the Hamiltonian matrix $H_{\alpha\beta}$ in every iteration in the self-consistent field (SCF) cycle.

The strength of the Kleinman-Bylander formalism lies generally in its scaling behavior. Rather than having to evaluate and store all $\bra{\phi_{\alpha}} \delta V_l \ket{\phi_{\beta}}$ matrix elements, resulting in $\mathcal{O}(N^2)$ scaling with the number of basis functions, the fully separable KB form only requires evaluation and storage of $N\times M$ projections with $M$ the (small) number of KB projector functions. In case of atom-centered basis sets the projections $\braket{\phi_{\alpha}^{(j)}}{\chi_{lm}^{(i)}}$ correspond furthermore to a two-center integral of basis function $\ket{\phi_{\alpha}^{(j)}}$ of atom $j$ at ${\bf R}_j$ and $\ket{\chi_{lm}^{(i)}}$ of PP $i$ at ${\bf R}_i$. The explicit expression to be evaluated is then:
\begin{align}
\label{formular:projection_full}
\nonumber
& \braket{\phi_{\alpha}^{(j)})}{\chi_{lm}^{(i)}} = \frac{1}{\int dr (\delta V_l(r) v_l(r) )^2} \nonumber \\ 
&  \int d{\bf r} \phi_{\alpha} ({\bf r}-{\bf R}_j) \frac{\delta V_l(|{\bf r}-{\bf R}_i|) v_l(|{\bf r}-{\bf R}_i|)}{|{\bf r}-{\bf R}_i|}Y_{lm}(\frac{{\bf r}-{\bf R}_i}{|{\bf r}-{\bf R}_i|}) \quad ,
\end{align}
which can be most efficiently computed with the help of spherical Bessel transformations\cite{Talman2003_1,Talman2003_2,Talman2009} already implemented in {\tt FHI-aims} \cite{Ren2012}. Again, the radial functions $v_l(r)$ and the potential $V_l(r)$, respectively potential differences $\delta V_l(r)$ of the projector are pretabulated in *.cpi format and provided as input for each relevant element. The total number of overlap integrals that needs to be computed at all is further reduced by locality, as overlap integrals are exactly zero if the distance of the involved atom centers exceeds a maximum value. This maximum value is given by the maximum extension of any basis function plus the maximum extension of any KB projector function, and is typically in the range of 8\,{\AA}.

For geometry optimization, molecular dynamics or vibrational analysis, force contributions on the PP with formal charge $q_i$ and centered at position ${\bf R}_i$ also need to be evaluated. Hellman-Feynman contributions arise from the embedding of the PP into the electrostatic fields of the electron density and all other nuclei in the QM region at positions ${\bf R}_j$ and with charges $Z_j$
\begin{eqnarray}
\nonumber
\boldsymbol{F}_i^{\rm loc} &=& - \sum_j^{n_{\rm atoms}} \frac{Z_j q_i}{|\boldsymbol{R}_j - \boldsymbol{R}_i|^3} (\boldsymbol{R}_j - \boldsymbol{R}_i) \\ &-& \int d^3 r \rho(\boldsymbol{r}) \boldsymbol{\nabla}_i V^{\rm loc}(\boldsymbol{r}) \quad .
\label{formular:local_force} 
\end{eqnarray}
Here, $\boldsymbol{\nabla}_i V^{\rm loc}(\boldsymbol{r})$ is the gradient of $V^{\rm loc}$ with respect to the position $\boldsymbol{R}_i$ and is needed on every global integration grid point. This is efficiently computed from a cubic spline interpolation of $V^{\rm loc}$ created before entering the SCF cycle. Analogous to the first term in eq. (\ref{formular:local_force}), the Coulomb interaction between nuclei and PPs gives rise to a force term of the same shape that complements the general QM-force expression for atoms in {\tt FHI-aims} (eqs. (69) and (71-75) in ref. \onlinecite{Blum2008}). As written and implemented, eq. (12) will work for non-periodic geometries. For periodic geometries, a formalism analogous to eq. (70) in ref. \onlinecite{Blum2008} would have to be adopted.

The overlap, eq. (\ref{formular:projection_full}), between any basis function and a KB-projector function $\ket{\chi_{lm}^{(i)}}$ of the PP $i$ at $\boldsymbol{R}_i$ is a function of $\boldsymbol{R}_i$. This gives rise to a force acting on the PP,
\begin{eqnarray}
\nonumber
&& \boldsymbol{F}^{\rm KB}_i = - \sum_k f_k \sum_{j, h}^{n_{\rm atoms}} \sum_{\alpha \in \mathcal{J}, \beta \in \mathcal{H} } c_{k \alpha} c_{k \beta}  \sum_{\chi \in \mathcal{I}} \sum_{lm} \nonumber\\
&& ((\frac{\partial}{\partial \boldsymbol{R}_i}\braket{\phi_{\alpha}^{(j)} }{\chi^{(i)}_{lm}}) E_l^{\rm KB}
\braket{\chi^{(i)}_{lm}}{\phi_{\beta}^{(h)}}\nonumber\\
&+& \braket{\phi_{\alpha}^{(j)}}{\chi^{(i)}_{lm}} E_l^{\rm KB} (\frac{\partial}{\partial \boldsymbol{R}_i} \braket{\chi^{(i)}_{lm}}{\phi_{\beta}^{(h)}}) )
\label{formular:nonlocal_force} \quad ,
\end{eqnarray}
where $f_k$ is the occupation number of Kohn-Sham state $k$, and $c_{k \alpha} c_{k \beta}$ is the density matrix.
$\mathcal{I}$,$\mathcal{J}$ and $\mathcal{H}$ are the subspaces of projector respectively basis functions belonging to PP $i$ and atoms $j$ and $h$, respectively. Again, derivatives of overlaps only have to be calculated for those pairs of functions, which have a non-zero overlap. Since the overlap, eq. (\ref{formular:projection_full}), is also a function of the position of atom $j$, this gives reciprocally rise to a negative force acting on atom $j$ complementing the general force expression on atoms.

In order to compute the Hartree potential, FHI-aims follows the strategy of an atom-centered multipole (MP) decomposition as introduced by Delley \cite{Delley1990}. Here, the density difference to a sum-over-free-atom density is partitioned into atom-centered components on the integration grid shells centered around individual atoms \cite{Blum2008}. Applying a MP decomposition to these atom-centered components leads (together with the known MPs of the free-atom density components) to the total MP components of the Hartree potential. \cite{Delley1990,Blum2008}. Although the ionic PPs do not introduce any electron density of their own, they do act as integration grid centers and are thus included into the partitioning. As a result, some electron density components are assigned to the PPs, even when no basis functions are centered at the position of the PP. The MP expansion always introduces an error $\rho(\boldsymbol{r}) - \rho_{\rm MP}(\boldsymbol{r})$ though, as the expansion is truncated beyond a maximum angular momentum 
$l_{\rm MP}^{\rm max}$. This error leads to a net force \cite{Blum2008},
\begin{equation}
\boldsymbol{F}_i^{\rm MP} = - \int d^3 r [\rho - \rho_{\rm MP}] \boldsymbol{\nabla}_i V^{\rm loc}(\boldsymbol{r}) 
\end{equation}
as the missing multipole terms move with the position of the center.

\section{Results}

\subsection{Reduction potentials in Fe--ZSM-5}

For a first demonstration of the established QM/MM embedded-cluster framework we address the zeolite ZSM-5 as an example 
for a semi-covalent system, where neither embedding PPs nor a polarizable MM environment are required. ZSM-5 is commonly used in catalytic applications \cite{UCL.171, UCL.172}, and has previously been the subject of investigations with the {\tt ChemShell} package coupled with the QM calculator {\tt Gamess-UK} \cite{UCL.23}. Belonging to the MFI structural family, ZSM-5 is a porous aluminium silicate framework (SiO$_{2}$ building units with Al-substitutions), which can be further doped with catalytically active transition metals such as Fe and Ti \cite{UCL.25}. Quantifying the reaction energetics at such transition metal centers is then an obvious milestone on the route to improved catalyst design. The accurate calculation of corresponding energies has proven problematic though, due to the difficulty in accounting for both the extended bulk framework and the electron localization around the 
active site. As semi-local DFT does not appropriately accomplish the latter ({\em vide infra}), calculations at hybrid xc level are at least required. Existing work for Fe--ZSM5 at this level of theory has hitherto been restricted to finite fragments that only reproduce the active site of the catalyst \cite{UCL.175, UCL.176, UCL.177, UCL.178, UCL.179}. As the surrounding framework undoubtedly does have an influence on the electrostatic potential at the active site \cite{Sherwood2003, UCL.185, UCL.186}, efficient QM/MM calculations are an ideal tool that in combination with powerful QM calculators such as {\tt FHI-aims} also allows to assess the performance of advanced functionals. Here, we illustrate this approach by calculating the reduction potential for intra-framework Fe$^{\mathrm{3+}}$ sites in MFI using semi-local, hybrid and double-hybrid functionals.
Double-hybrid functionals such as XYG3 \cite{XYG3} are still rather uncommon in the application to extended systems, while their implicit ability to improve electron localisation could render them particularly appealing for systems such as Fe--ZSM-5.

\begin{figure}
\includegraphics[width=8.5 cm]{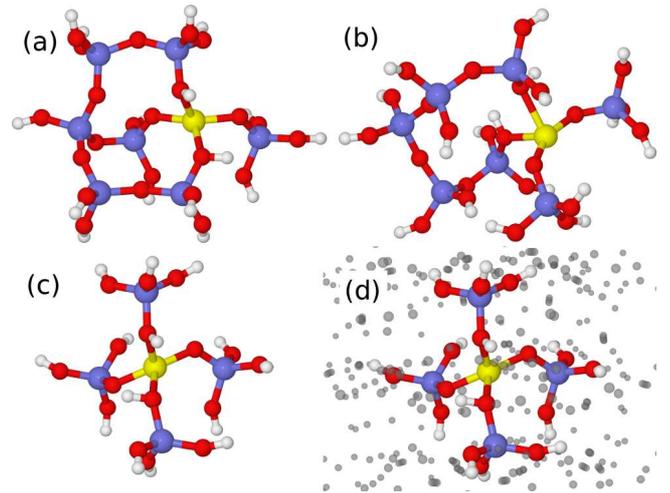}
\caption{Illustration of the QM clusters used in our calculations of Fe-ZSM-5: (a) M7-T1; (b) Z6-T4; and (c) I2-T12. Fe, Si, O and H atoms are represented in yellow, mauve, red and white, respectively. The extended embedding environment included in the QM calculation of (c) is illustrated in (d), with the MM point charges represented in grey.}
\label{fig2}
\end{figure}

We construct three different embedded clusters from the undoped parent MFI framework, each centered on a tetrahedral Si atom (T-sites) of differing structural nature within a framework active site \cite{Sherwood2003}: a T-site at a straight channel (M7-T1), at a sinusoidal channel (Z6-T4) and at a channel intersection (I2-T12), cf. Fig. \ref{fig2}. Around this central atom the QM region includes all other members of the active site, as well as any extra Si atoms that are first neighbours to the active site, and all the corresponding linking oxygen atoms. All incomplete bonds from the oxygen atoms are saturated with H link-atoms, and a bond-dipole correction is added to the MM representation \cite{UCL.186} . Concentric regions of MM atoms are then added; first an active region up to a radius of 10.58\,{\AA} (20 Bohr) and then a fixed region up to a radius of 21.17\,{\AA} (40 Bohr). This results in a total number of atoms in the embedded clusters in the range 2155 -- 2180. The MM interactions are represented 
using the forcefield of Hill and Sauer \cite{UCL.102}, and energy/
force calculations are performed using the {\tt DL\_POLY} package. The accuracy of the {\tt FHI-aims}--{\tt ChemShell} coupling is validated by 
calculating the same embedded clusters also with {\tt Gamess-UK} as QM calculator, and using the same TZVP Gaussian basis sets of Ahlrichs and Taylor \cite{UCL.101} in both QM codes. At the PBE \cite{107} xc level of theory and using converged integration grid settings in both codes differences in total energies are below 18\,meV for each cluster, while forces on QM atoms differ by less than 0.05\,eV/{\AA}. 

With this confidence, production calculations with {\tt FHI-aims} are henceforth performed with the code's own numerical basis sets. For the three embedded clusters at PBE level, already tier1 leads hereby to total energies lower by 1\,eV than for the TZVP basis set, demonstrating the better description of e.g. the near-nuclear potential and kinetic energy integrals by the purposely designed NAO basis sets. The Fe$^{3+}$ active center is introduced to the embedded clusters by replacing the central Si atom. The reduced Fe$^{\mathrm{2+}}$ state is described by coordinating an additional H atom to an O atom directly adjacent to the central Fe species. The reduction potential can then be defined as
\begin{align}
\nonumber
& E^{\text{Red}}(\text{Fe$^{3+/2+}$}) = \\ 
& E^{\text{tot}}(\text{Fe$^{2+}$-MFI}) + \frac{1}{2}E^{\text{tot}}(\text{H$_{2}$}) - E^{\text{tot}}(\text{Fe$^{3+}$-MFI}),
\end{align}
where $E^{\mathrm{tot}}(\mathrm{H}_{2}$) is the energy of a gas-phase hydrogen dimer. Calculations are performed for the semi-local PBE and BLYP, for the hybrid PBE0 and B3LYP, as well as for the double-hybrid XYG3 functionals \cite{107,PBE0,UCL.180, UCL.181, UCL.182, UCL.183, XYG3}. All geometries are fully relaxed at the corresponding semi-local level, with hybrid and double-hybrid calculations performed on these geometries, and specifically those of the according correlation treatment (i.e. B3LYP and XYG3 on the BLYP geometry, PBE0 on the PBE geometry). The XYG3 calculations are furthermore performed post-SCF on the optimized B3LYP Kohn-Sham orbitals. At tier2 basis set level the obtained reduction potentials are already numerically converged to within 1 meV for the semi-local and hybrid functionals. At the double hybrid-level a sufficient convergence to within $\pm 20$\,meV can instead only be reached using the tier3 basis set for Fe and for all other (light) species the NAO-VCC-4Z basis set that 
specifically converges the unoccupied-space sums \cite{IgorZhang2013}. The more diffuse functions contained in this basis set then require an ensuing counterpoise correction \cite{BoysBernardi1970} though.

\begin{table}
\begin{tabular}{lccc}
\hline
& M7-T1 & Z6-T4 & I2-T12 \\
\hline
\\
$E^{\mathrm{Red}}_{\mathrm{PBE}}$(Fe$^{3+/2+}$)  & -0.18 & -0.16 & -0.18 \\
$E^{\mathrm{Red}}_{\mathrm{BLYP}}$(Fe$^{3+/2+}$) & -0.29 & -0.39 & -0.17 \\ \hline
$E^{\mathrm{Red}}_{\mathrm{PBE0}}$(Fe$^{3+/2+}$) & -0.05 &  0.04 & -0.03 \\
$E^{\mathrm{Red}}_{\mathrm{B3LYP}}$(Fe$^{3+/2+}$)& -0.09 &  0.01 & -0.02 \\ \hline
$E^{\mathrm{Red}}_{\mathrm{XYG3}}$(Fe$^{3+/2+}$) & -0.34 & -0.20 & -0.18 \\
\\
\hline
\end{tabular}
\caption{\label{table1} Computed reduction potentials (in eV) for Fe$^{3+/2+}$ embedded intra-framework within an MFI siliceous structure. Active sites are modeled as the M7-T1, Z6-T4 and I2-T12 sites shown in Fig. \ref{fig2}. Calculations performed using semi-local and hybrid functionals used the tier2 basis level for all atoms; at the double hybrid-level the NAO-VCC-4Z basis set was used for all atoms except Fe, for which a tier3 basis set was used.}
\end{table}

Table \ref{table1} compiles the obtained reduction potentials for Fe at the three structural sites in the MFI framework. At all levels of theory the differences between these three sites are very small, indicating a low structure sensitivity for the catalytic properties of the active center. This structure insensitivity warrants to directly compare to previous experimental voltammetry by P\'{e}rez-Ram\'{i}rez \textit{et al.}, which gave very small reduction energies for intra-framework Fe$^{3+}$ of the order 0.05\,eV $\gtrsim V_{\mathrm{Red}}$(Fe$^{3+/2+}$) $\gtrsim -0.15$\,eV \cite{UCL.187}. There is, thus, a huge effect induced by the siliceous framework, when considering that the corresponding reduction potential of Fe$^{3+}$ in a hexagonally coordinated [Fe(H$_{2}$O)$_{6}$]$^{3+}$ complex was for instance calculated as 1.07\,eV at PBE level \cite{UCL.189}. This framework effect is well captured by essentially all xc levels studied, which in good overall agreement with the experimental findings yield 
small reduction potentials scattering around zero. More specifically, the majority of the xc-treatments agrees that the reduction potential is negative, indicating stability for Fe(II) in the material. 
Interestingly, the double-hybrid functional XYG3 yields reduction potentials closer to the ones at the semi-local (BLYP) level, and thus makes up for most of the change observed at the hybrid (B3LYP) functional level. At least within the small scatter of Table \ref{table1} there is thus no clear trend in the calculated reduction potentials that would prominently reflect the allegedly gradually increasing electron localization achieved in the sequence BLYP $\rightarrow$ B3LYP $\rightarrow$ XYG3.

\subsection{Water splitting reaction energetics at TiO$_2$ (110)}

As a show case application to a polarizable material, where both effective core potentials at the QM/MM boundary and core-shell potentials in the MM region are essential, we next consider the TiO$_2$(110) surface. Among many other applications, this semiconductor catalyst is particularly known for its ability to oxidize water using light \cite{Diebold2003,Fujishima2008,Henderson2011}. The desire for a molecular-level understanding of this intriguing property motivates detailed mechanistic studies unraveling the  chemical steps involved. In surface science 
studies especially, the rutile TiO$_2$(110) surface has been frequently studied \cite{Diebold2003,Yates2005}, with recent theoretical work considering a reaction mechanisms including an OOH intermediate on a defect-free surface \cite{Norskov2004,ValdesRossmeisl2008}. Characteristic for the field of surface catalysis, these calculations were performed at the semi-local DFT level, specifically with the rPBE \cite{rPBE} functional. Arguably because of its 
acceptable description of hydrogen bonds, this functional has been frequently employed in the context of water dissociation and photocatalysis at TiO$_2$ \cite{Zhang2005,Kroes2006,ValdesRossmeisl2008,83,Monica2011,Monica2013}, but, of course, as a GGA functional it still suffers from the well-known electron delocalization problems. The possibility to 
perform hybrid and double-hybrid level calculations efficiently with our QM/MM setup offers therefore an ideal platform to assess how much this affects the reaction energetics.

\begin{figure}
\includegraphics[width=8.5 cm]{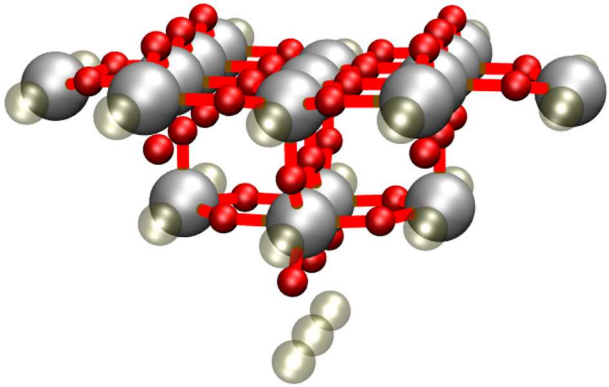}
\caption{Perspective view of the Ti$_{17}$O$_{34}$ cluster, exposing at its top center the fivefold coordinated Ti adsorption site of the TiO$_2$(110) surface. Ti atoms are shown as large white spheres, O atoms as small red spheres, and semi-transparent grey spheres mark the positions where PPs represent the immediately surrounding Ti-cations.}
\label{fig3}
\end{figure}

For the QM region and systematic size convergence tests we employ a series of three clusters proposed by Anmal and Heyden \cite{Heyden2010}, Ti$_{17}$O$_{34}$, Ti$_{29}$O$_{58}$ and Ti$_{33}$O$_{66}$, with the Ti$_{17}$O$_{34}$ cluster including the employed PPs at the boundary region illustrated in Fig. \ref{fig3}.  At each level of theory the embedded clusters are constructed on the basis of optimized rPBE lattice positions. Formal charges are used for the embedding point charges (+4 for Ti and -2 for oxygen), whereby any positive point charge in a 8\,{\AA} vicinity of the QM region was replaced by Ti$^{4+}$ PPs. The MM region extends up to a distance of 27\,{\AA} from the central adsorption site, involving a total number of 4480 point charges. The interactions in the MM region are described with the well-established polarizable TiO$_2$ forcefield from ref. \onlinecite{Catlow1993}, where only the  spring potential between oxygen shell and oxygen core has been modified to $V = k d^2 (cosh(\frac{r}{d}) -1 ) 
$ with $k = 20$\,eV/{\AA}$^2$ and $d = 0.1$\,{\AA}. This improves the representation of the dielectric constant of the MM region when setting it up with the DFT lattice parameters. In order to 
describe the intrinsic surface polarization properly, the position of the top-row O MM atoms is fully optimized for the clean TiO$_2$(110) surface. For the neutral systems studied here, further geometry optimization and self-consistent polarization of the MM region in response to the adsorbates is neglected. The position of the adsorbates on top of the central fivefold-coordinated Ti site in the QM region, cf. Fig. \ref{fig3}, is optimized at the rPBE level; higher-level calculations are then performed for these fixed geometries. As reference we also compute supercell geometries with a 5 O-Ti$_2$O$_2$-O trilayer slab, a 50\,{\AA} vacuum, and applying the identical surface geometry optimization protocol as for the embedded clusters. For $(1 \times 2)$ and $(2 \times 4)$ surface unit cells  $(4 \times 4 \times 1)$ and $(
2 \times 2 \times 1)$ Monkhorst-Pack k-point grids were used, respectively, while for the density of states (DOS) calculations of the clean $(1 \times 2)$ cell this grid was increased to $(40 \times 40 \times 1)$.

Following earlier theoretical work \cite{Norskov2004,ValdesRossmeisl2008,Oberhofer2013} we assume the water oxidation pathway at defect-free TiO$_2$(110) to proceed along four electron-coupled proton transfer steps:
\begin{subequations}
	\label{eq:nr-orig}
	\begin{align} \nonumber
		 \quad & {\rm H_2O} + (^*) \rightarrow {\rm OH^*} + {\rm H^+} + e^- \label{eq:nr-stepa}\\ \nonumber
		 \quad & {\rm OH^*} \rightarrow {\rm O^*} + {\rm H^+} + e^- \\ \nonumber
		 \quad & {\rm H_2O} + {\rm O^*} \rightarrow {\rm OOH^*} + {\rm H^+} + e^- \\ \nonumber
		 \quad & {\rm OOH^*} \rightarrow {\rm O_2} + (^*) + {\rm H^+} + e^- \quad , 
	\end{align}
\end{subequations}
where the asterisk stands for the five-fold coordinated Ti centers offered by the catalytic surface $(^*)$ and particles attached to them (e.g.~O$^*$), respectively. Central energetic quantities for this pathway are correspondingly the binding energies of O, OH and OOH, defined as
\begin{equation}
E_{\rm b}[X] \;=\; E^{\rm tot}[X{\rm @TiO_2}] - E^{\rm tot}[{\rm TiO_2}] - E^{\rm tot}[X] \quad .
\label{total_energy_QMMM}
\end{equation}  
Here, $E^{\rm tot}[X{\rm @TiO_2}]$ and $E^{\rm tot}[{\rm TiO_2}]$ are the total energies of the TiO$_2$(110) surface with and without adsorbate $X$, respectively, and $E^{\rm tot}[X]$ is the total energy of the isolated adsorbate. At the hybrid and double-hybrid functional levels, identical spin states are obtained in corresponding QM/MM and slab calculations. At the semi-local level, which significantly suffers from self-interaction, 
this needed to be explicitly ensured by fixing the spin states to those of the higher-level calculations, i.e. doublet for OH and OOH, and triplet for O. This is an important issue, as most semi-local functionals yield the wrong spin-polarization for the O adsorbate, and already due to this produce a large deviation in the corresponding binding energy with respect to the higher-rung calculations \cite{Oberhofer2013}. Similar to the findings made for the zeolitic system, tier2 basis sets readily converge all $E_{\rm b}$ to within 10\,meV at the semi-local and hybrid level, while at the double-hybrid level the 
valence-correlation consistent NAO-VCC-4Z basis set for all O atoms together with a counterpoise correction \cite{BoysBernardi1970} was required to reach this level of convergence.

\begin{table}
\begin{tabular}{lccc}
\hline
$E_{\rm b}$       &   O  & OH &  OOH\\
\hline
PBC \\
(2$\times$1)      &-0.46&-0.53&-0.17\\ 
(4$\times$2)      &-0.46&-0.52&-0.26\\
\hline
free cluster \\
Ti$_{17}$O$_{34}$ &-0.52&-0.51& 0.01\\ 
\hline
QM/MM \\
Ti$_{17}$O$_{34}$ &-0.45&-0.49&-0.26\\ 
Ti$_{29}$O$_{58}$ &-0.45&-0.50&-0.25\\
Ti$_{33}$O$_{66}$ &-0.44&-0.50&-0.27\\
\hline
\end{tabular}
\caption{\label{table2} Calculated rPBE binding energies (in eV) of O, OH and OOH on the rutile TiO$_2$(110) surface. Compared are results from different clusters with and without embedding against results obtained from a periodic supercell setup and differing surface unit-cells.}
\end{table}

Table \ref{table2} compares the rPBE binding energies computed with the embedded clusters against results obtained for a bare cluster and from supercell calculations with two coverages, 0.5\,monolayer (ML) in a $(2 \times 1)$ surface unit-cell and 0.25\,ML in a $(4 \times 2)$ surface unit-cell. In the QM/MM setup, convergence with respect to the size of the QM region is rapidly reached. Already the Ti$_{17}$O$_{34}$ cluster yields binding energies converged to within 10\,meV, which we attribute to the correct polarization treatment of our solid-state embedding approach. In contrast, in the supercell calculations large $(4 \times 2)$ surface unit-cells are needed to reach the low-coverage limit modeled in the QM/MM approach for the OOH adsorbate. This, in turn, we attribute to the high dipole moment of this adsorbate, which sensitively feels nearby periodic images -- and equally the incorrect electrostatic potential of a bare cluster geometry. For all three adsorbates, representing a wide 
range of electron 
affinities and ionization potentials, low-coverage supercell and size-converged QM/MM calculations agree within 20\,meV, demonstrating the accuracy of the established solid-state embedding approach and its huge potential for future work on explicitly charged systems (without need for intricate charge-compensation schemes).

\begin{figure}
\includegraphics[width=8.5 cm]{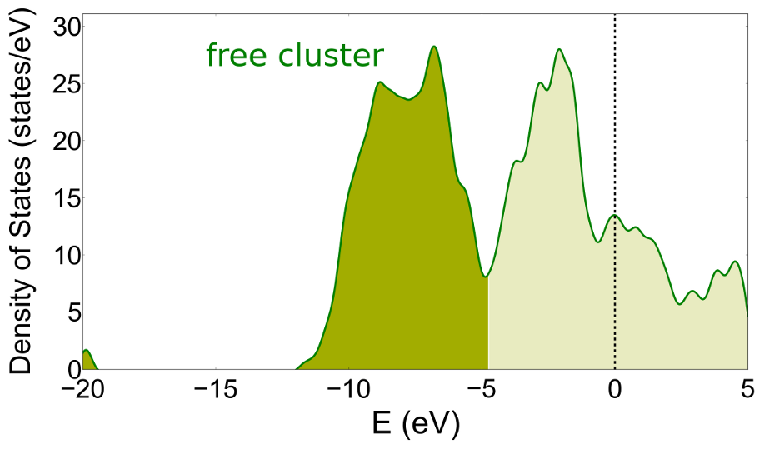}
\includegraphics[width=8.5 cm]{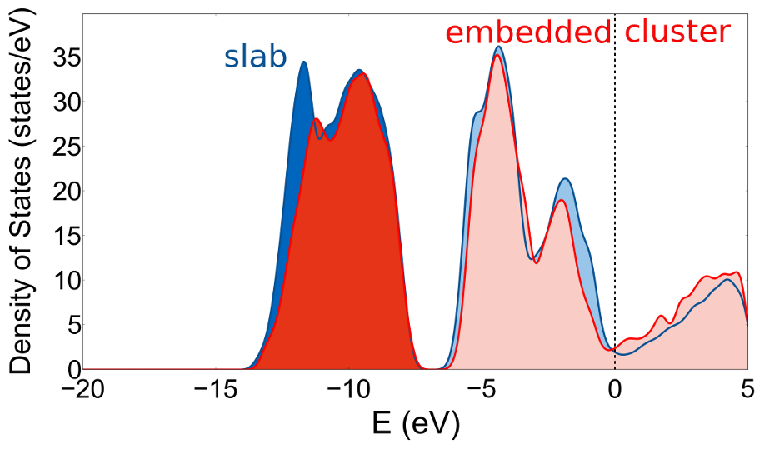}
\caption{Comparison of the total density of states (DOS) for the clean TiO$_2$(110) surface at the rPBE level: Even a relatively large free Ti$_{54}$O$_{108}$ cluster (upper panel) still shows huge deviations with respect to the corresponding embedded Ti$_{54}$O$_{108}$ cluster (lower panel), whereas the latter achieves almost perfect agreement with the corresponding PBC supercell calculations (lower panel), in particular in the valence and lower conduction band region. The vacuum level represents the zero reference throughout, and filled states are depicted in darker colour.}
\label{fig4}
\end{figure}

Most importantly, this agreement between supercell and QM/MM calculations is not only restricted to the
\textit{binding energetics, but extends for instance also to the more sensitive underlying electronic structure.}
This is illustrated in Fig. \ref{fig4}, which compares the total DOS for the clean TiO$_2$(110) surface. Whereas the DOS of an already quite large free Ti$_{54}$O$_{108}$ cluster still exhibits huge deviations with respect to the DOS obtained for the same-size embedded cluster, there is essentially perfect agreement between the latter and the DOS obtained from a PBC supercell calculation. This agreement extends not only to the band gap i.e. 
relative difference of highest occupied molecular orbital (HOMO) and lowest unoccupied molecular orbital (LUMO), but also to the absolute band edge positions, which are, of course, of utmost importance for photoelectrochemical applications like the light-driven water oxidation reaction studied here. The position of the delocalized LUMO is hereby more sensitive 
to the finite QM-region of the solid-state embedding approach. For the smaller Ti$_{17}$O$_{34}$ cluster it is 0.3\,eV higher in energy compared to its converged position in the large Ti$_{54}$O$_{108}$ cluster, whereas the position of the localized HOMO is essentially independent of the employed cluster size.

\begin{table}
\begin{tabular}{lccc}
\hline
$E_{\rm b}$    &  O  &  OH & OOH \\
\hline
rPBE           &-0.45&-0.49&-0.26\\
BLYP           &-0.60&-0.62&-0.30\\
PBE            &-0.71&-0.79&-0.43\\ \hline
B3LYP          &-0.36&-0.56&-0.35\\
PBE0           &-0.39&-0.64&-0.41\\ 
HSE06          &-0.41&-0.66&-0.43\\ \hline
XYG3           &-0.36&-0.64&-0.53\\
\hline
\end{tabular}
\caption{
\label{table3} Calculated binding energies (in eV) of O, OH and OOH on the rutile TiO$_2$(110) surface. Results are obtained from an embedded Ti$_{17}$O$_{34}$ cluster at different levels of theory (GGA, hybrid and double-hybrid), see text.}
\end{table}

Table \ref{table3} compiles the results obtained at GGA, hybrid and double-hybrid level. Generally, we observe a somewhat larger scatter among the three GGAs tested as compared to the group of hybrid and double-hybrid functionals. As widely perceived, in particular the PBE functional seems to be more on the overbinding side. Interestingly, hybrid and double-hybrid xc level reduces the bond strength of the O adsorbate and increases the bond strength of the OOH intermediate as compared to the semi-local description. Together with its overall weaker binding, rPBE mimicks this best, which to some extent seems to support the arguments made in its favor in preceding work at the semi-local level. The inclusion of PT2 correlation in the XYG3 double-hybrid functional primarily changes the most polarization-affected OOH binding as compared to its parent B3LYP hybrid functional. This breaks the trend in bond strength OH $>$ O $>$ OOH that was consistently obtained for the semi-local and hybrid functionals. 

\begin{table}
\begin{tabular}{lccc}
\hline
$E_{\rm b}^{\rm SHE}$    &  O  &  OH & OOH \\
\hline
rPBE           &4.80&2.43&4.60\\
BLYP           &4.75&2.34&4.59\\
PBE            &4.90&2.34&4.61\\ \hline
B3LYP          &4.86&2.37&4.78\\
PBE0           &4.92&2.36&4.86\\ 
HSE06          &4.87&2.32&4.84\\ \hline
XYG3           &4.88&2.39&4.89\\
\hline
\end{tabular}
\caption{
\label{table4} Calculated binding energies (in eV) of O, OH and OOH on the rutile TiO$_2$(110) surface, but referenced against the standard hydrogen electrode (SHE) following eqs. (10-12) of ref. \onlinecite{Norskov2011}. Results are obtained from the same computational setup and settings as in Table \ref{table3}.}
\end{table}

Notwithstanding, these differences are primarily due to the description of the bare radical in the gas phase, which is used as reference in the definition of $E_{\rm b}$ in eq. (\ref{total_energy_QMMM}) and which is, of course, most sensitively affected by a varying degree of electron localization achieved at the different xc-levels. To circumvent this, adsorption energies in photoelectrochemical calculations are often given with respect to the computational standard hydrogen electrode (SHE) \cite{Norskov2004}. Table \ref{table4} reproduces the same binding energies of Table \ref{table3} with respect to this reference and demonstrates that this indeed removes most of the scatter and leads to consistent trends at all xc functional levels studied. Under the above described constraint that the GGA functionals are enforced to yield the correct spin polarization, these results thus suggest that despite the known electron localization problems, a description at the semi-local 
level seems indeed sufficient for 
computational screening work relying on trends and correlations between reaction intermediate binding energies\cite{Norskov2011,Koper2010}, rather than quantitative differences within ~0.2\,eV \cite{Norskov2011}. Of course, this may be quite different when addressing adsorption at defects, in charged states, or when calculating reaction barriers -- with the present QM/MM-setup then forming an ideal tool to conduct higher-level calculations either directly for production or as reference.

\section{Summary and Conclusions}

We have presented a general-purpose QM/MM-setup integrating the all-electron electronic structure theory code {\tt FHI-aims} into the {\tt ChemShell} package. In this course, pseudopotential functionality was implemented into {\tt FHI-aims}, which in the solid-state embedding context is employed to prevent spurious charge leakage out of the QM zone into nearby cationic MM charges. The {\tt FHI-aims} NAO basis sets enable particularly efficient access to exact exchange and second order perturbation theory. The here established QM/MM-approach is therefore ideally suited for hybrid and double-hybrid DFT calculations of extended systems.

The performance of our approach was demonstrated by the application to two different examples:
The calculation of the Fe reduction potential in a Fe-substituted ZSM-5 zeolitic framework and the calculation of adsorption energies of reaction intermediates in the water oxidation at defect-free TiO$_2$(110). In both cases our results confirm the crucial importance of the appropriate description of the long-range electrostatics achieved through the solid-state embedding. Systematic comparison to PBC supercell calculations for the more demanding TiO$_2$ system even shows that the possibility to adequately capture its polarizability through core-shell potentials in the MM region yields reliable absolute band edge positions, which predestines the approach for (photo-)electrochemical applications.

The calculation of the reduction potentials and adsorption energies at hybrid and double-hybrid xc level demonstrates the value of being able to scrutinize more readily available semi-local DFT data. In case of the reduction potentials the higher-level reference data confirms the low structure sensitivity and the huge effect of the siliceous framework, with the small remaining scatter increasing the confidence in the quantitative numbers. In case of the adsorption energies the higher-level reference data reveals that the electron delocalization problem of the semi-local description can lead to wrong spin polarizations, which in turn would lead to qualitatively wrong results. If the correct spin polarization is enforced, the benchmark against the hybrid and double-hybrid numbers indicates that the semi-local binding energetics of the neutral reaction intermediates is reliable enough for computational screening work, in particular if referencing is done against the standard hydrogen electrode rather than 
delicate bare radicals. Whether this prevails for more demanding cases like adsorption at (charged) defects remains to be seen and is the topic of on-going work -- with the here established solid-state embedding approach providing a powerful tool that enables such calculations at higher xc-level efficiently and without the need for intricate charge compensation schemes.

\section{Acknowledgements}

We thank J\"urgen Wieferink (now Aleo Solar) for help with his logarithmic spherical Bessel transform infrastructure in {\tt FHI-aims}, Igor Zhang for his advice regarding numerical polarisation basis sets, and Thomas W. Keal for discussions regarding the implementation of the {\tt FHI-aims}/{\tt ChemShell} interface. AJL, CRAC, PS and AAS acknowledge funding from EPSRC grant EP/IO30662/1, MRF and CRAC acknowledge funding from EPSRC grant EP/I03014X/1, and DB acknowledges a scholarship from the TUM International Graduate School of Science and Engineering. Further support came from the Solar Technologies Go Hybrid initiative of the State of Bavaria. The authors acknowledge the use of the following high-performance computing facilities, and associated support services: HECToR, via our membership of the UK HPC Materials Chemistry Consortium (EP/L000202); IRIDIS, provided by the Centre for Innovation; HYDRA, provided by the Max-Planck Supercomputing Center. Parts of this work were carried out under the HPC-
EUROPA2 
project (project number: 228398), with the support of the European Community - Research Infrastructure Action of the FP7.

\section{Appendix: Full PP-functionality in {\tt FHI-aims}}

In the QM/MM-context the PP-functionality is exclusively used to prevent overpolarization of the QM charge due to immediately adjacent MM cationic monopoles. As such, the monopole term of the cation is replaced by a bare ionic PP with equal formal charge. In contrast, when using the PP in regular electronic structure calculations, valence electrons would be considered for the pseudoized atom and would then contribute to the total electronic energy. As the xc energy and potential are not additive with respect to the electron density, a non-linear core correction (NLCC) \cite{Louie1982,Porezag1999} may be required to make up for the missing core density of the pseudoized atom and therewith allow to properly capture the xc contribution of the added valence electrons. The essential idea of the NLCC is to simply add the core density $\rho_{\rm core}$ of a free atom in the calculation of the pertinent parts of the total energy expression.

{\tt FHI-aims} calculates the total energy, referred to as $E^{\rm QM}$ in the main text, as \cite{Blum2008}
\begin{eqnarray}
\nonumber
 E^{\rm QM} &=& \sum_i^{N_{\rm states}} f_i \epsilon_{i} -  \int d r^3 [\rho(\boldsymbol{r})v_{\rm xc}[ \rho](\boldsymbol{r})]
 \\ \nonumber& +&  E_{\rm xc}[\rho] 
 \\&-&  \frac{1}{2}\int d r^3 [\rho(\boldsymbol{r})V_{\rm H}[\rho](\boldsymbol{r})]  + E_{\rm nuc-nuc}
\label{equation:harris}
\end{eqnarray}
with $f_i$ the occupation number of Kohn-Sham state $\epsilon_i$ and $V_{\rm H}$ the Hartree potential. The eigenvalues 
$\epsilon_i$ are hereby obtained in the standard manner from diagonalization of the Hamiltonian of the Kohn-Sham system, $\hat{h}_{\rm KS} = \hat{t}_{\rm s} + \hat{v}_{\rm ex} + \hat{v}_{\rm es} + \hat{v}_{\rm xc}$. Here, $\hat{t}_{\rm s}$ is the kinetic energy operator, $\hat{v}_{\rm ex}$ the external potential, $\hat{v}_{\rm es}$ the electrostatic potential, and $\hat{v}_{\rm xc}$ the xc potential for the single electrons. The NLCC correspondingly affects the first three terms in eq. (\ref{equation:harris}), which are thus replaced as follows:
\begin{eqnarray}
\nonumber 
E_{\rm xc}^{\rm NLCC}[\rho + \rho_{\rm core}] &=& \sum_i^{N_{\rm states}} f_i \bra{\psi_i} \hat{v}_{\rm xc}[\rho + \rho_{\rm core}]\ket{\psi_i} \\ \nonumber
&-& \int d^3 r \; \rho \; V_{\rm xc}[\rho + \rho_{\rm core}] \\
&+& \int d^3 r \; (\rho + \rho_{\rm core}) \epsilon_{\rm xc}[\rho + \rho_{\rm core}] \quad .
\label{equation:e_nlcc}
\end{eqnarray}
Even though not of concern for the QM/MM focus, this expression was also implemented into {\tt FHI-aims}. Thus, if an "empty" site (no nucleus, but including basis functions and integration grids) is placed atop a PP in {\tt FHI-aims}, our present implementation also allows to perform norm-conserving PP calculations in {\tt FHI-aims}. As smoothness of the core density is not an issue for NAO basis sets and integration grids designed for an all-electron code, we hereby employ the full atomic core density $\rho_{\rm core}$ and not a smoothed auxiliary representation as is commonly done in plane-wave codes \cite{Fuchs1999}.

As the added core density is atom centered, a Pulay-type force term on the pseudoized atom $i$ arises when adding the NLCC \cite{Pulay1969,Gerratt1968I,Gerratt1968II}. Differentiating eq. (\ref{equation:e_nlcc}) determines this term as
\begin{eqnarray}
\nonumber
\boldsymbol{F}^{\rm NLCC}_i &=& - \frac{\partial E_{\rm xc} ^{\rm NLCC}[\rho]}{\partial \boldsymbol{R}_i} 
 \nonumber\\ &=& - \int  \frac{\delta ((\rho + \rho_{\rm core}) \epsilon_{\rm xc}[\rho + \rho_{\rm core}])}{\delta (\rho + \rho_{\rm core})}  \frac{\partial \rho_{\rm core}}{\partial \boldsymbol{R}_i}   
  \nonumber\\ &=& - \int d r^3 V_{\rm xc}[\rho + \rho_{\rm core}]  \frac{\partial \rho_{\rm core}}{\partial \boldsymbol{R}_i} \quad   
\end{eqnarray}
at the level of LDA.


\end{document}